# Franson-type experiment realizes two-qubit quantum logic gate

Kaoru Sanaka, Karin Kawahara, and Takahiro Kuga

*Institute of Physics, University of Tokyo at Komaba, 3-8-1 Komaba, Meguro-ku, Tokyo 153-8902, Japan*

Quantum computers promise great improvements in solving problems such as factoring large integers, simulating quantum systems, and database searching[1-3]. Using a photon as a quantum bit (qubit) is one of the most promising ways to realize a universal quantum computer because the coherent superposition state of a photon is very robust against various sources of decoherence. However, it is too difficult to realize two-qubit (photon) gates because it requires huge nonlinearity between photons[4,5]. Here we show the realization of a controlled-NOT (CNOT) gate, the most important and elemental two-qubit gate for quantum computation, by extending our previous research[6]. The heart of our experiment is the conditional measurement of two-photon coincidences in the Franson-type experiment[7]. The photon counting measurement plays the same role as the nonlinearity required for the two-qubit gate, and our system reproduces the truth table of the CNOT gate. Furthermore, we create an entangled state from the superposition state by our gate, which is clear evidence that our gate works as a quantum logic gate. Our results make it possible to manipulate the quantum state of photons including entanglement and represent significant progress in the operation of various algorithms in quantum computation.

Any universal quantum logic circuit can be created by using a set of single-qubit gates and two-qubit CNOT gates[8,9]. A CNOT gate is used for the operation between two qubits; the transformations of the gate can be described by $|0\rangle_1|0\rangle_2 \to |0\rangle_1|0\rangle_2$, $|0\rangle_1|1\rangle_2 \to |0\rangle_1|1\rangle_2$, $|1\rangle_1|0\rangle_2 \to |1\rangle_1|1\rangle_2$, and $|1\rangle_1|1\rangle_2 \to |1\rangle_1|0\rangle_2$, where $|x\rangle_1$, (x=0,1) refers to the control qubit and $|x\rangle_2$, (x=0,1) refers to the target qubit (Fig. 1(a)). Optical implementation of the CNOT gate is quite difficult to realize, unlike single-qubit gates. The conditional phase shift between photons (qubits) has been demonstrated by a cavity-QED experiment[10]. However, the amount of phase shift is less than 180°, which is not sufficient to realize a CNOT gate. Some efficient schemes have been proposed to produce the conditional phase shift of 180° by using linear optics and multi-photon entanglement[11-13]. However, the efficiency in generating multi-photon entanglement is so low that one can hardly utilize such schemes in quantum computation[14,15].

Figure 1(b) shows the optical implementation of a CNOT gate using interferometers 1 and 2. Here, H and V represent horizontal and vertical polarization (coding for qubit $|0\rangle$ and $|1\rangle$). Our CNOT gate is constructed by a setup similar to the Franson-type experiment[7]. The photon in interferometer 1 acts as the control qubit, and the photon in interferometer 2 acts as the target qubit in the CNOT gate.

Interferometer 1 is constructed by two polarizing beam splitters (PBS). The injected photon is split into the short or long path of the interferometer depending on its polarization state at the first PBS, and combined again in the same path at the second PBS. In Fig. 1(b), $\alpha|H\rangle_1 + \beta|V\rangle_1$ ($|\alpha|^2+|\beta|^2=1$) represents an arbitrary polarization state of a photon in interferometer 1. The length of the long (short) path of the interferometer is L (S) and the path difference is L-S. The injected photon of horizontal polarization ($\alpha=1$ and $\beta=0$) exits the interferometer earlier than the photon of vertical polarization ($\alpha=0$ and $\beta=1$). The time difference of the outputs is (L-S)/c.

Interferometer 2 is constructed by two 50%-50% non-polarizing beam splitters (BS) and a half-wave plate (HWP) that rotates the polarization state by 90°. The injected photon whose polarization state is $|H\rangle_2$ ($|V\rangle_2$) is split into the short or long path with the probability of 1/2. If the photon takes the short path, the state of the photon does not change, $|H\rangle_2$ ($|V\rangle_2$). However, if the photon takes the long path, the polarization state is transformed to $|V\rangle_2$ ($|H\rangle_2$) due to the HWP. The split photon is combined again in the same path by the second BS. The output photon whose polarization state is the same as the input state $|H\rangle_2$ ($|V\rangle_2$) exits the interferometer earlier than the photon whose polarization state is transformed to $|V\rangle_2$ ($|H\rangle_2$). Because the path difference is also L-S, the time difference of the outputs is the same as in interferometer 1.

We injected simultaneously generated photon pairs into interferometers 1 and 2. If the states of the injected photon pairs are $\alpha|H\rangle_1 + \beta|V\rangle_1$ and $|H\rangle_2$ ($|V\rangle_2$), the positions and polarized states of output photon pairs are shown in Fig. 1(b). We measure the coincidence counts of photon pairs with the time window ($\Delta T$) and set the time window smaller than the arrival time difference (L-S)/c. Under this condition, the output states of photon pairs that contribute to the coincidence counts are $1/2|H\rangle_1|H\rangle_2$, $1/2|H\rangle_1|V\rangle_2$, $1/2 e^{i(\theta_1+\theta_2)}|V\rangle_1|V\rangle_2$, and $1/2 e^{i(\theta_1+\theta_2)}|V\rangle_1|H\rangle_2$ if the input states of photon pairs are $|H\rangle_1|H\rangle_2$, $|H\rangle_1|V\rangle_2$, $|V\rangle_1|H\rangle_2$, and $|V\rangle_1|V\rangle_2$, respectively. In these equations, $\theta_1$ ($\theta_2$) is the phase difference caused by the path difference of interferometer 1 (2). We can choose the phases to satisfy $e^{i(\theta_1+\theta_2)}=1$. The transformation of the states with our system is then the same as for the CNOT gate, except for the probability factor of 1/4. The transformation probability is listed in Table 1(a).

An experimental demonstration of the CNOT gate is represented schematically in Fig. 2. We use a CW beam (420 nm, 1.4 mW) from a violet laser diode (NICHIA-NLHV500A) as a pump source to generate photon pairs. The violet light is sent to the waveguide-type nonlinear device[16], and horizontally polarized photon pairs around 840-nm wavelength are generated in the process of spontaneous Type-I parametric down-conversion (PDC) with high efficiency[6]. Dichroic mirrors are used to separate the violet beam from the photon pairs.

Generated photon pairs are separated using a BS and injected into interferometers 1 and 2 after passing through half-wave plates (HWP1 and HWP2). HWP1 and HWP2 are used to prepare the state of input qubits $|H\rangle_1|H\rangle_2$, $|H\rangle_1|V\rangle_2$, $|V\rangle_1|H\rangle_2$, and $|V\rangle_1|V\rangle_2$. Interferometer 1 is composed of a PBS and two retroreflectors (dotted line). Interferometer 2 is composed of a BS, a half-wave plate (HWP), and two retroreflectors (solid line). They work exactly as our proposed scheme of the CNOT gate shown in Fig. 1(b). Path differences of both interferometers are set to about 56 cm and correspond to an arrival time difference of photons of 1.9 ns at the detector.

The correlation of the polarization of output photon pairs from the interferometers is measured by using polarization analyzers consisting of HWP3, HWP4 and two PBSs. After passing through the analyzers, photon pairs are directed onto single-photon detectors $D_1$ and $D_2$ (EG&G SPCM-AQR14). The signal from $D_1$ is used for the start signal of a time-to-amplitude converter, and the signal from $D_2$ is used for the stop signal after it passes through an electrical delay line. We record the coincidence counts for 10 seconds with the time window $\Delta T = 1.0$ ns $< (L-S)/c$ under computer control. The experimental results are listed in Table 1(b). These results are obtained by dividing the coincidence counts of a specific outcome by the sum of all possible outcomes. We have very good agreement between calculations and experimental results. Our system thus reproduces the truth table of the CNOT gate.

We also generate the entangled state from superposition state ($\alpha = \beta = 1/\sqrt{2}$ in Fig.1(b)) using our method. In this case, HWP1 in Fig. 2 is set to rotate the linearly polarized state of photons by 45°, while HWP2 is set to maintain the polarization state. The input state of photon pairs to our system $|\Psi_{IN}\rangle$ is

$$|\Psi_{IN}\rangle = \frac{1}{\sqrt{2}}(|H\rangle_1 + |V\rangle_1)|H\rangle_2 . \qquad (1)$$

The output state of photon pairs to our system $|\Psi_{OUT}\rangle$ becomes

$$|\Psi_{OUT}\rangle = \frac{1}{2\sqrt{2}}(|H\rangle_1|H\rangle_2 + e^{i\theta_2}|H\rangle_1|V\rangle_2 + e^{i\theta_1}|V\rangle_1|H\rangle_2 + e^{i(\theta_1+\theta_2)}|V\rangle_1|V\rangle_2) . \qquad (2)$$

Under the condition $\Delta T < (L-S)/c$, the state that contributes to the coincidence counts can be described as

$$|\Psi_{OUT}\rangle \xrightarrow{\Delta T < (L-S)/c} |\Psi_{entangle}\rangle = \frac{1}{2\sqrt{2}}(|H\rangle_1|H\rangle_2 + e^{i(\theta_1+\theta_2)}|V\rangle_1|V\rangle_2) . \qquad (3)$$

The state of equation (3) is called the two-photon polarization entangled state with the conditional measurement of $\Delta T < (L-S)/c$ and is referred to as time entanglement in Franson-type experiments[7,17-19]. The transformation to the entangled state is achieved with the probability of 1/4 for input photon pairs because only one-half of the photons input to interferometer 2 can arrive at $D_2$ and a half of these photons can contribute to the conditional measurement.

We first verify that the polarization states of photon pairs that contribute to the coincidence counts are only $|H\rangle_1|H\rangle_2$ and $|V\rangle_1|V\rangle_2$ when the time window of coincidence counter is set $\Delta T = 1.0$ ns. This is done by comparing the detection probabilities of all four possible polarization combinations, HH, HV, VH, and VV. We accumulate the coincidence counts for 20 seconds. The results are shown in Fig. 3(a). The probability of expected combinations of HH and VV are $P_{HH} = 0.44 \pm 0.03$ and $P_{VV} = 0.41 \pm 0.02$.

Showing the existence of $|H\rangle_1|H\rangle_2$ and $|V\rangle_1|V\rangle_2$ states alone is a necessary but not sufficient experimental criterion for verifying polarization entanglement. We have to demonstrate that the output states from our system are indeed a coherent superposition of two states. We set HWP3 and HWP4 to rotate the polarization state by 45° to measure the two-photon interference caused by the polarization entanglement. A piezoelectric ceramic actuator (PZT) can move the long path of interferometers ($1V = 69 \pm 7$ nm) and make it possible to manipulate the phase $\theta = \theta_1 + \theta_2$ in equation (3). The interferometers are stabilized by using a reference laser.

Observed coincidences for 20 seconds as a function of the applied voltage to the PZT (path difference) are shown in Fig. 3(b). The visibility of the two-photon interference is $V = 0.44 \pm 0.16$. The quantum entanglement is characterized by the fidelity $F = (P_{HH} + P_{VV} + V)/2$ [20]. By substituting these experimental data, we obtain the fidelity $F = 0.65 \pm 0.10$. The fact that $F$ is sufficiently larger than 0.5 ensures that the polarization entangled state of photon pairs is generated.

Quantum circuits using more than two CNOT gates are also realized by extending our method. In this case, the path difference of interferometers used in the ($i+1$)-th CNOT gate ($\Delta L_{i+1}$, $i = 1, 2, \cdots$) should be twice as long as that used in the previous ($i$-th) CNOT gate ($\Delta L_i$). If this condition is not satisfied, the arrival time difference of photon pairs after several stages of CNOT gates can be smaller than the time window of the coincidence measurement, and these photon pairs generate undesired coincidence counts. Because the time resolution of the photon detector is less than 100 ps in the current technology, quantum circuits using a few CNOT gates can be realized with interferometers whose path lengths are around a few meters.

We realize the CNOT gate by using the conditional measurement of two-photon coincidences in the Franson-type experiment. We also generate the two-photon polarization entanglement by using our method and verify that its fidelity is $F = 0.65 \pm 0.10$. Multi-qubit operation with our CNOT gate is also possible if the simultaneous multi-photon generators are available[21-23]. The representation of n qubits requires $2^n$-optical paths using linear optics and single-photon sources[24,25]. However, our method requires only n paths to represent n qubits and does not cause the difficulty of exponential growth in the number of optical paths with the number of qubits.   Our CNOT gate thus has a major advantage in constructing a universal quantum computer of multi-qubit operation.
---------------

*Acknowledgements*

We are grateful to the members of the Optical Disk Systems Development Center at Matsushita Electric Industrial Co. for their experimental cooperation.   This work was supported by Matsuo Foundation and the Research Foundation for Opto-Science and Technology.

**a**     Input HWP Settings

<table>
<tr><th rowspan="2">Output HWP Settings</th><th></th><th>H H</th><th>H V</th><th>V H</th><th>V V</th></tr>
</table>

| Output HWP Settings | H H | H V | V H | V V |
|---|---|---|---|---|
| H H | 0.25 | 0 | 0 | 0 |
| H V | 0 | 0.25 | 0 | 0 |
| V H | 0 | 0 | 0 | 0.25 |
| V V | 0 | 0 | 0.25 | 0 |

**b**     Input HWP Settings

| Output HWP Settings | H H | H V | V H | V V |
|---|---|---|---|---|
| H H | 0.244 (6) | 0.004 (0) | 0.002 (1) | 0.001 (0) |
| H V | 0.002 (0) | 0.243 (3) | 0.000 (0) | 0.004 (1) |
| V H | 0.001 (1) | 0.002 (1) | 0.006 (2) | 0.236(10) |
| V V | 0.002 (0) | 0.001 (0) | 0.241(11) | 0.009 (1) |

Table 1. (a) The transformation probability of polarization states of photon pairs with our system. The operation of CNOT gate can be realized with the probability of 0.25. (b) Experimental results. These results are obtained by dividing the coincidence counts of a specific outcome by the sum of all possible outcomes ( $X(Y) = X \pm Y \times 10^{-3}$ ).

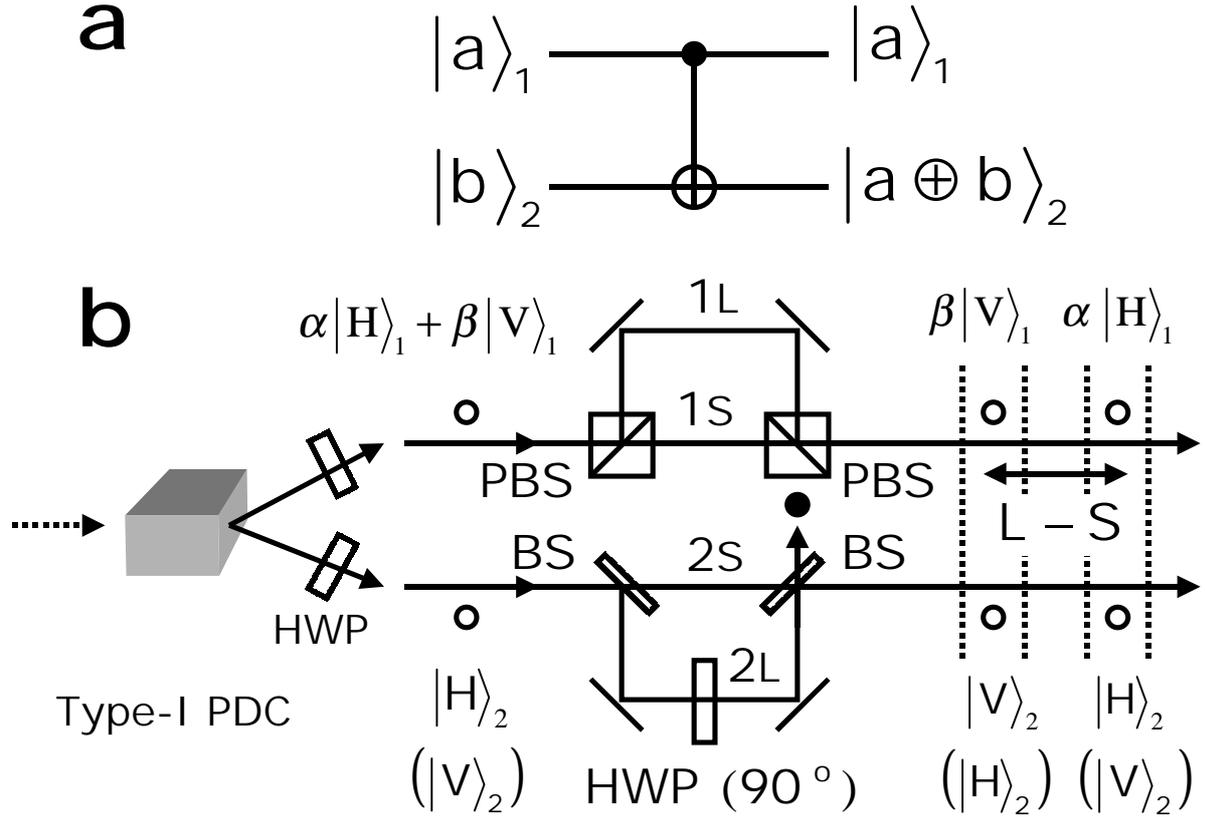

Figure 1. (a) The representation of controlled-NOT (CNOT) gate. $|a\rangle_1$ refers to control qubit and $|b\rangle_2$ refers to the target qubit. (b) Optical implementation of CNOT gate by using our method. Our CNOT gate is constructed using a setup similar to the Franson-type experiment. The photon in interferometer 1 acts as the control qubit and the photon in interferometer 2 as the target qubit in the CNOT gate. The positions and polarized states of output photon pairs are shown at the right side of the figure when the simultaneously generated photon pairs are injected into interferometers 1 and 2.

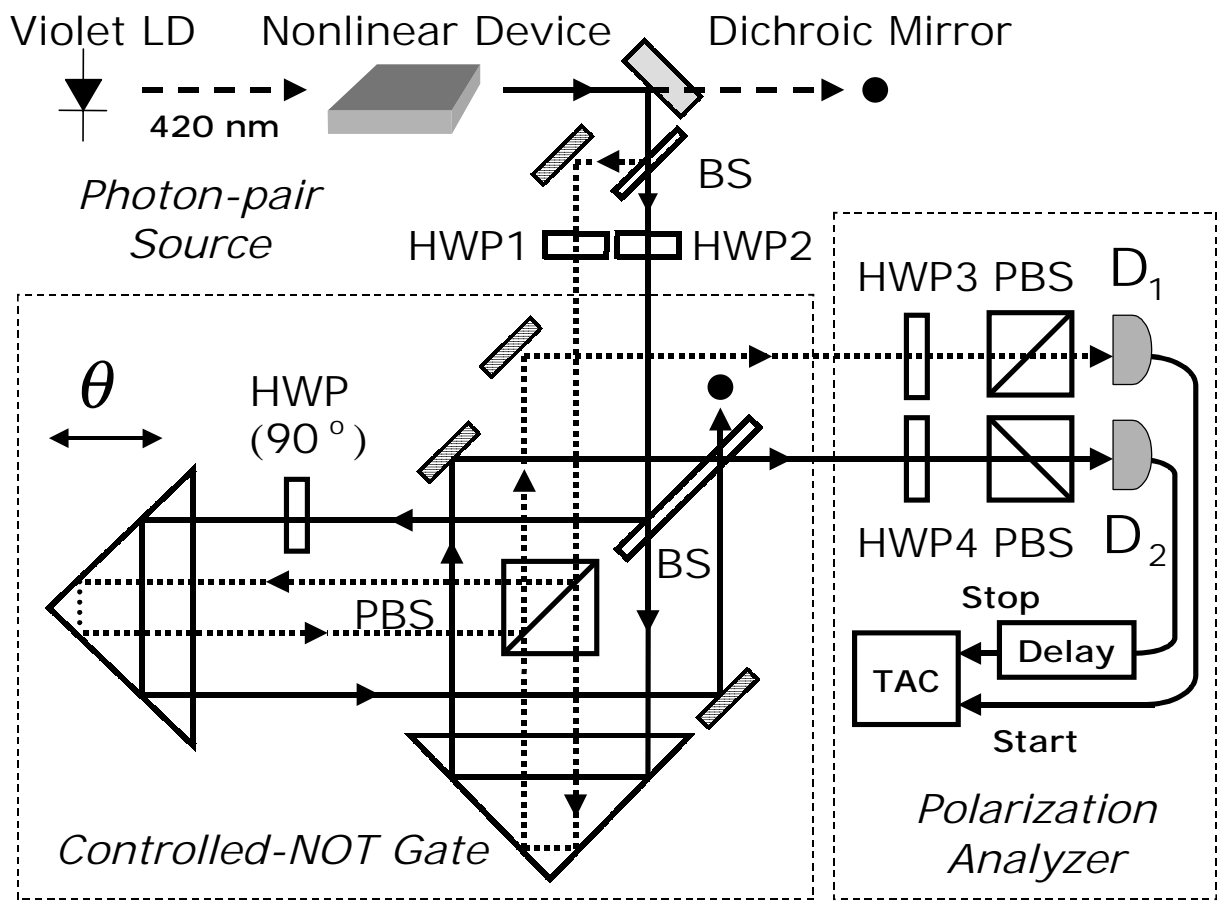

Figure 2. Experimental setup to demonstrate the CNOT gate. Pairs of horizontally polarized photons are generated by type-I parametric down-conversion in the waveguide-type nonlinear device pumped by a CW laser beam (420 nm, 1.4 mW). Dichroic mirrors are used to separate the violet laser beam from photon pairs. By using a BS, generated photon pairs are separated and injected into interferometers 1 and 2 after passing through half-wave plates (HWP). HWP1 and HWP2 are used to prepare the state of input qubits. Interferometer 1 (dotted line) is composed of a PBS and two retro-reflectors. Interferometer 2 (solid line) is composed of a BS, a half-wave plate (HWP), and two retro-reflectors. Path differences of both interferometers are set to about 56 cm and correspond to an arrival time difference of photons of 1.9 ns at the detector. The correlation of the polarization of photon pairs output from the interferometers is measured by using polarization analyzers (HWP3, HWP4 and two PBS's) and a coincidence counter (single photon detectors $D_1$ and $D_2$ and TAC).

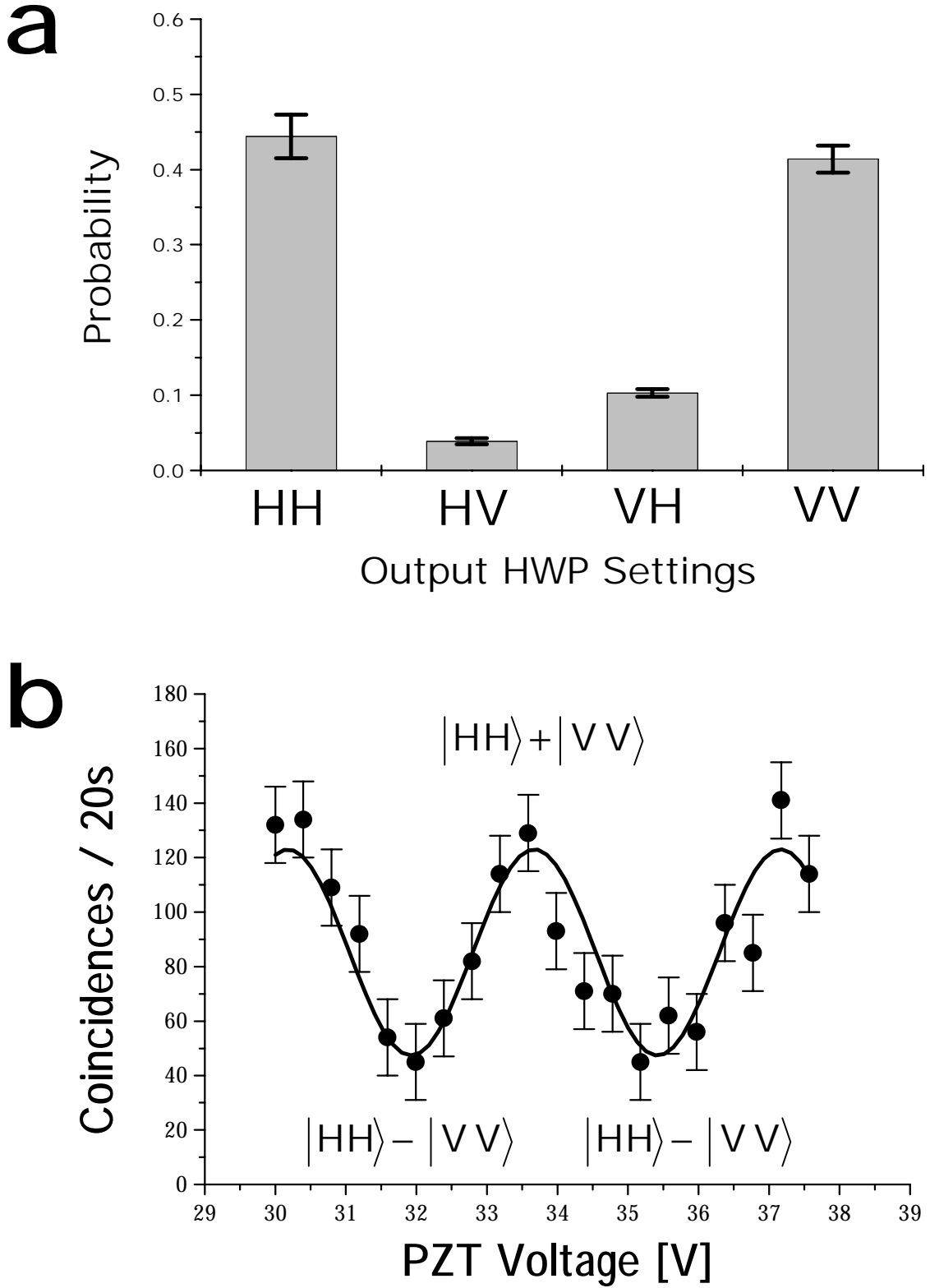

Figure 3. (a) The observed detection probability of the photon pairs in polarization entanglement. The probability of combinations of HH and VV are $P_{HH} = 0.44 \pm 0.03$ and $P_{VV} = 0.41 \pm 0.02$, while the probabilities of other combinations are much less than $P_{HH}$ and $P_{VV}$. (b) Two-photon interference measurement caused by the polarization entanglement. The dots are observed coincidences for 20 seconds at various voltages applied to the PZT. The solid curve is the best fit and obtained visibility is $V = 0.44 \pm 0.16$.